\documentstyle[prb,aps,preprint]{revtex}

\begin{document}

\draft


\title{Quasiparticle Interface States in Junctions Involving d-Wave
Superconductors}

\author{Yu.~S.~Barash}
\address{P.N. Lebedev Physical Institute, Leninsky Prospect 53, Moscow 117924,
Russia\\
and\\ Department of Physics, \AA bo Akademi, Porthansgatan 3, FIN-20500 \AA bo,
Finland
}




\maketitle


\begin{abstract}
Influence of surface pair breaking, barrier transmission and
phase difference on quasiparticle bound states in junctions
with $d$-wave superconductors is examined. Based on the
quasiclassical theory of superconductivity, an approach is
developed to handle interface bound states. It is shown in SIS'
junctions that low energy bound states get their energies
reduced by surface pair breaking, which can be taken into
account by introducing an effective order parameter for each
superconductor at the junction barrier. More interestingly, for
the interface bound states near the continuous spectrum the
effect of surface pair breaking may result in a splitting of
the bound states. In the tunneling limit this can lead
to a square root dependence of a nonequilibrium Josephson
current on the barrier transmision, which means an enhancement
as compared to the conventional critical current linear in the
transmission. Reduced broadening of bound states in NIS
junctions due to surface pair breaking is found.
\end{abstract}





\section{Introduction}
The important role in the Josephson effect of quasiparticle bound states
localized in contacts between isotropic $s$-wave superconductors, is well known
from studies of superconductor-normal metal-superconductor (SNS) junctions and
quantum point contacts. Even for spatially constant order parameters of equal
moduli on the two sides of a junction (no surface pair breaking), Andreev
reflection takes place on account of the phase shift between them.
Current-carrying interface bound states with phase-dependent energies
are formed at the junction\cite{fur90,fur191,been191,been291}. In short
symmetric junctions, these bound states are known to carry all the Josephson
current, while in asymmetric junctions the contribution from the continuous
spectrum is of importance as well.

The situation becomes more complicated in junctions involving $d$-wave
superconductors where a quasiparticle, depending on its momentum, can see both
substantial modulus distortions or (and) a sign change in the order parameter
in a reflection or a transmission process at a junction. Several kinds of bound
states occur in this case, each with a different dependence on the
quasiparticle momentum according to the incoming and outgoing quasipaticle
trajectories as well as the crystalline orientations on the two sides of the
junction. In particular, low energy interface bound states are of interest,
associated with changes of sign of the order parameter in reflection or
transmission events. They become dispersionless zero-energy states both in the
limiting case of zero transmission (impenetrable wall) or (and) in the opposite
limit of a ballistic junction\cite{hu,yhu,tan196,rb98}. In certain conditions,
zero-energy (or low-energy) bound states can result in an anomalous low
temperature behavior of the Josephson critical current\cite{tan296,bbr96} and
in characteristic peaks and jumps in the I-V curves\cite{bs}. In normal
metal-insulator-superconductor (NIS) junctions they lead, at sufficiently low
temperatures, to a zero-bias conductance peak\cite{hu,tan95,buch95,fog197,bbs}.

As incoming quasiparticles with a momentum along the interface normal see the
same $d$-wave order parameter as the outgoing ones, the bound states they
occupy are to some extent analogous  to those in junctions with isotropic
$s$-wave superconductors. These bound states can dominate the charge transport
across a junction for certain orientations of d-wave superconductors if the
transmission coefficient is sufficiently selective, limiting the transport of
current to quasiparticles with momentum directions close to the interface
normal, as is always the case for thick junctions. The difference between
$s$-wave and $d$-wave cases, in general, is present even for the momentum
orientation parallel to the interface normal being associated with surface pair
breaking substantially more pronounced in anisotropically paired
superconductors. As compared to isotropic $s$-wave superconductors, boundary
conditions for anisotropic order parameters are very different because of
surface pair breaking even within the Ginzburg-Landau
theory\cite{amb74,bgz,samokh,bgs}.  Depletion of the modulus of the order
parameter in the vicinity of a contact modifies the bound state energies seen
in non-selfconsistent models with constant absolute values of the order
parameters. Besides, additional bound states can appear for quasiparticles in
an effective potential well formed by the spatially dependent moduli of order
parameters on the two sides of the junction\cite{buch95,bs,bbs}. Since surface
pair breaking depends upon crystal-to-interface orientations, it can strongly
modify corresponding angular dependences of bound state energies and, in
particular, the critical current obtained for a non-selfconsistent spatially
constant order parameters\cite{bs99}.

Assuming order parameters with spatially constant moduli on both sides of a
junction, calculations of the energies of current-carrying quasiparticle bound
states can be found in the literature admitting the presence of a phase shift
and any value of the transmission\cite{fur90,fur191,been291,tan196,rb98}. The
effect of surface pair breaking combined with a phase shift and the influence
of a finite transmission on the interface bound states have, however, to the
best of my knowledge, not been studied yet. Below, in Section II, I shall
develop an analytical approach for studying the combined effects of surface
pair breaking, a phase shift and the transmission coefficient on bound states,
localized, in particular, at a contact between two $d$-wave superconductors.
The approach is based on the quasiclassical theory of superconductivity focused
on the problem of interface bound states. A key point of the consideration is
that retarded propagators at the interface take quite large values (pole-like
terms) at energies $\omega$ close to an interface bound state energy
$\varepsilon_B(\bbox{p}_f)$. Expanding propagators in powers of
$(\omega-\varepsilon_B(\bbox{p}_f))$ one can introduce, in a first
approximation, an ansatz for the bound states, which reduces the Eilenberger
equations and the normalization condition to one scalar differential equation
for a quasiclassical phase $\eta$ to be completed with asymptotic and boundary
conditions for its solutions. This equation was derived earlier in Ref.\
\onlinecite{bbs} for the particular case of impenetrable boundary and real
order parameter. Actually the equation has a more general character, as it is
associated with the approach where the Eilenberger equations transform to a
scalar Riccati equation\cite{schop}. It has the same form as that obtained on
the basis of the Bogoliubov-de Gennes equations within the WKB
approximation\cite{bard69,been191}, being applicable both to descrete and to
continuous spectrum. Boundary conditions formulated below for the
quasiclassical equation are new. It is remarkable, that the boundary condition
for the quasiclassical quantity $\eta$, in accordance as it is with Zaitsev's
boundary conditions, can be formulated separately from all other quantities,
just like the one-scalar boundary condition for the equation for $\eta$
mentioned above. It is a great deal simpler as compared to Zaitsev's original
form and admits analytical results. The same boundary condition for $\eta$ can
be derived also starting from the boundary conditions for the Andreev equations
irrespective of whether a descrete or continuous spectrum is considered.

On this basis I study interface bound states in both SIS' (see Section III) and
NIS (see Section IV) junctions. It is shown for SIS' junctions that surface
pair breaking results in reducing the energies of low energy bound states which
can be taken into account by introducing an effective order parameter for each
superconductor at the junction barrier. For interface bound states near the
edge of the continuos spectrum, the effect of surface pair breaking turns out
to be more interesting, resulting in splitting the bound state energies. In the
tunneling limit this can lead to a square root dependence of the nonequilibrium
Josephson current on the barrier transmision, which means an enhancement as
compared to the conventional critical current linear in transmission. For NIS
junctions the influence of surface pair breaking on broadening the bound state
is considered.

\section{Quasiclassical Theory of Interface Bound States}

\subsection{Ansatz for bound states}

Quasiclassical theory of superconductivity is based on Eilenberger's equations
for the quasiclassical matrix propagator. In the case of a clean singlet
anisotropically paired superconductor the equations for the retarded propagator
$\hat{g}^R$ reduce to the following $2\times2$ matrix form:
\begin{eqnarray}
\left[\varepsilon\hat{\tau}_3-\hat{\Delta}(\bbox{p}_f,\bbox{r}),
\hat{g}^R(\bbox{p}_f,\bbox{r};\varepsilon)\right] +
i \bbox{\rm v}_f\hspace*{-2pt} \cdot \hspace*{-2pt}
\bbox{\nabla} \hspace*{-2pt}_{\bbox{r}}\:
\hat{g}^R(\bbox{p}_f,\bbox{r};\varepsilon)=0 \enspace, \label{eil} \\
\big[\hat{g}^R(\bbox{p}_f,\bbox{r};\varepsilon)\big]^2=
-\pi^2\hat{1} \enspace.
 \qquad\qquad\qquad
\label{norm}
\end{eqnarray}
Here, $\varepsilon$, $\bbox{p}_f$, $\bbox{\rm v}_f$ and $\hat\Delta$ are the
quasiparticle energy, the momentum at the Fermi surface, the Fermi velocity and
the order parameter matrix respectively.  A ``hat'' indicates matrices in Nambu
space and $\hat\tau_3$ is a Pauli matrix in this space. The propagator $\hat g$
and the order parameter matrix $\hat\Delta$ have the form
\begin{eqnarray}
\hat g=\left(\begin{array}{cr}
 g   &  f \\
 f^+ & -g
\end{array}\right)\enspace
\qquad \mbox{and} \qquad
\hat \Delta=\left(\begin{array}{cc}
 0     &  \Delta \\
-\Delta^* &     0
\end{array}\right)\enspace.
\label{matrix}
\end{eqnarray}
Henceforth the superscript $R$ is dropped for simplicity.  The boundary
conditions for the quasiparticle propagators at a smooth interface with
transmission $D(\bbox{p}_{f})=1-R(\bbox{p}_{f}) \ (R(\bbox{p}_{f})$ the
reflectivity coefficient of the interface) are given by Zaitsev's relations
(see Refs.\ \onlinecite{zai,kur}) which can be written in the following matrix
form
\begin{eqnarray}
\hat{d}_l\ \hat{s}_l^2=i\alpha\ \left [\hat{s}_l,\ \hat{s}_r\left (\pi-
\frac{i}{2}\hat{d}_l\right )\right ] \enspace , \label{zai1} \\
\hat{d}_l=\hat{d}_r  \enspace ,
\label{zai2}
\qquad\qquad\qquad
\end{eqnarray}
with $\alpha=(1-R)/(1+R)$, $\hat s_{l(r)}=\hat g_{l(r)}(\bbox{p}_{f,l(r)})+
\hat g_{l(r)}(\underline{\bbox{p}}_{f,l(r)})$, $\hat d_{l}=\hat
g_{l}(\bbox{p}_{f,l})-\hat g_{l}(\underline{\bbox{p}}_{f,l})$, $\hat
d_{r}=-\hat g_{r}(\bbox{p}_{f,l})+\hat g_{r}(\underline{\bbox{p}}_{f,l})$, and
the propagators are taken on the left or the right side of the interface.
Equations (\ref{zai1}), (\ref{zai2}) connect, at the interface, the propagators
of an incoming quasiparticle from the left and the right sides of the
interface with momenta $\bbox{p}_{f,l}$ , $\bbox{p}_{f,r}$  and the propagators
of the reflected quasiparticles with the momenta $\underline{\bbox{p}}_{f,l}$,
$\underline{\bbox{p}}_{f,r}$. For specular reflection, the momentum parallel to
the interface is conserved, i.e.,
$\bbox{p}^\parallel_{f,l}=\underline{\bbox{p}}_{f,l}^\parallel
=\bbox{p}^{\parallel}_{r,f}=\underline{\bbox{p}}^{\parallel}_{f,r}$.  For a
complete determination of the quasiclassical propagator one has to take into
account that deep inside the superconductor the propagator approaches its bulk
value.

In the presence of a quasiparticle bound state with the energy
$\varepsilon_B(\bbox{p}_f)$, manifesting dispersion dependence on the Fermi
momentum $\bbox{p}_f$, the quasiclassical propagator $\hat g$ has a pole at
$\varepsilon=\varepsilon_B(\bbox{p}_f)$.  Following Ref.\ \onlinecite{bbs},
one can introduce the residue of the propagator $\hat g$
\begin{equation}
\hat{\tilde g}(\bbox{p}_f,\bbox{r};\varepsilon_B(\bbox{p}_f))=
\lim_{\varepsilon\to\varepsilon_B(\bbox{p}_f)}
\left[(\varepsilon-\varepsilon_B(\bbox{p}_f))
\hat g(\bbox{p}_f,\bbox{r};\varepsilon)\right]
\enspace ,
\label{gtilde}
\end{equation}
which is finite, satisfies the same transport equation (\ref{eil}) as $\hat g$,
but completed with the relation
\begin{equation}
\big[
\hat{\tilde g}(\bbox{p}_f,\bbox{r};
\varepsilon_B(\bbox{p}_f))
\big]^2=0
\enspace ,
\label{mbc}
\end{equation}
rather than the normalization condition (\ref{norm}).

Linear boundary relations (\ref{zai2}) being applied to
$\widehat{\widetilde{d}}$, remain unchanged. At the same time nonlinear
boundary conditions (\ref{zai1}), taken at a bound state energy, simplify in a
significant fashion. Terms containing multiplications of three propagators
dominate in (\ref{zai1}), if each propagator is well described by a large
pole-like term. Other terms in Eq.(\ref{zai1}), with only two propagators, can
be neglected under conditions in question. Then Eq.(\ref{zai1}) reduces to
\begin{equation}
\hat{\tilde s}_l^2-\frac{\displaystyle\alpha}{\displaystyle 2}
\{\hat{\tilde s}_l,\hat{\tilde s}_r\}=0
\enspace .
\label{red}
\end{equation}
As the left hand side of matrix equation (\ref{red}) is proportional to the
unit matrix, it leads to one independent scalar equation only.

Eilenberger's equations for $\hat{\tilde g}$ can be solved in terms of the
following ansatz:
\begin{eqnarray}
 \tilde{f}^{+}(\bbox{ p}_{f},x;\varepsilon_B(\bbox{p}_f))=
\tilde g  (\bbox{ p}_{f},x;\varepsilon_B(\bbox{p}_f))
exp(-i\eta(\bbox{p}_f, x))
\enspace , \nonumber \\
\tilde{f}(\bbox{ p}_{f},x;\varepsilon_B(\bbox{p}_f))=
-\tilde g  (\bbox{ p}_{f},x;\varepsilon_B(\bbox{p}_f))
exp(i\eta(\bbox{p}_f, x))
\enspace .
\label{symmetry}
\end{eqnarray}
This ansatz was introduced earlier in Ref.\ \onlinecite{bbs} for the particular
case of a real order parameter and impenetrable wall. In general, the
substitution (\ref{symmetry}), satisfying Eq.\ (\ref{mbc}), allows for the
quantity $\eta$ (as well as $\tilde g(\bbox{ p}_{f},x;
\varepsilon_B(\bbox{p}_f))$ and $\varepsilon_B(\bbox{p}_f)$) to take complex
values $\eta=\eta'+i\eta''$
($\varepsilon_B(\bbox{p}_f)=\varepsilon'_B(\bbox{p}_f)+
i\varepsilon''_B(\bbox{p}_f)$). Complex values of $\eta$ and
$\varepsilon_B(\bbox{p}_f)$ imply, in particular, broadened quasiparticle bound
states (due to finite quasiparticle lifetime) discussed in the last section of
the present article for NIS-junctions on account of finite transmission.

Introducing the phase of the complex order parameter $\Delta(\bbox{p}_f,x)=
|\Delta(\bbox{p}_f,x)|e^{i\phi(\bbox{p}_f,x)}$, one obtains from (\ref{eil}),
(\ref{mbc}) with substitution (\ref{symmetry}):
\begin{equation}
\tilde g(\bbox{p}_f,x;\varepsilon_B(\bbox{p}_f))=
\tilde g_0(\bbox{p}_f,\varepsilon_B(\bbox{p}_f))
\exp\left(-\displaystyle\frac{\displaystyle 2}{{\rm v}_{f,x}}
\int\limits^x_0
|\Delta(\bbox{p}_f,\tilde x)|\sin\left(\eta(\bbox{p}_f,\tilde x)-
\phi(\bbox{p}_f,\tilde x)\right)
d\tilde x\right) \enspace ,
\label{gx}
\end{equation}
together with the following equation for $\eta$:
\begin{equation}
-\frac{{\rm v}_{f,x}}{2}\partial_x\eta(\bbox{p}_f,x)+
\varepsilon_B(\bbox{p}_f)-
|\Delta(\bbox{p}_f,x)|\cos\left(\eta(\bbox{p}_f,x)-
\phi(\bbox{p}_f,x)\right)=0 \enspace .
\label{pe}
\end{equation}
According to Eq.(\ref{gx}), under some conditions the residue $\hat{\tilde g}$
vanishes exponentially in the bulk of the superconductor. Then, the quantity
$\hat{\tilde g}$ describes a quasiparticle state bound to the interface.
Furthermore, equation (\ref{pe}) coincides with one obtained many years ago on
the basis of WKB approximation for Bogoliubov-de Gennes
equations\cite{bard69,been191}, being applicable both to descrete and to
continuous spectrum. For continuous spectrum $\eta(\bbox{p}_f,\tilde x)-
\phi(\bbox{p}_f,\tilde x)$ should be a purely imaginary quantity in accordance
with Eq.(\ref{gx}). The same equation (\ref{pe}) is known to appear as well
within the approach transforming the Eilenberger equations to a
scalar Riccati equation\cite{schop}.

The asymptotic condition for $\eta$, which garantees solution (\ref{gx}) to
vanish in the bulk, takes in the right half space
($x\rightarrow+\infty$) the form
\begin{equation}
{\rm v}^r_{f,x}(\bbox{p}_{f,r})
\sin\left(\eta'_{r,\infty}(\bbox{p}_{f,r})-\phi_{r,\infty}(\bbox{p}_{f,r})
\right) >0
\enspace ,
\label{asr}
\end{equation}
while in the limit $x\rightarrow-\infty$
\begin{equation}
{\rm v}^l_{f,x}(\bbox{p}_{f,l})
\sin\left(\eta'_{l,\infty}(\bbox{p}_{f,l})-\phi_{l,\infty}(\bbox{p}_{f,l})
\right) <0
\enspace .
\label{asl}
\end{equation}

Substitution (\ref{symmetry}) essentially simplifies relations (\ref{zai2}),
(\ref{red}) , resulting, in particular, in the following boundary condition for
equation (\ref{pe})
$$
R\sin\left(\frac{\displaystyle \eta_{l,0}(\bbox{p}_{f,l})-
\eta_{l,0}(\underline{\bbox{p}}_{f,l})}{\displaystyle 2}\right)
\sin\left(\frac{\displaystyle\eta_{r,0}(\bbox{p}_{f,r})-
\eta_{r,0}(\underline{\bbox{p}}_{f,r})}{\displaystyle 2}\right)=
\qquad\qquad\qquad\qquad\qquad\qquad
$$
\begin{equation}
\qquad\qquad\qquad\qquad\qquad\qquad\qquad
D\sin\left(\frac{\displaystyle \eta_{l,0}(\bbox{p}_{f,l})-
\eta_{r,0}(\underline{\bbox{p}}_{f,r})}{\displaystyle 2}\right)
\sin\left(\frac{\displaystyle\eta_{l,0}(\underline{\bbox{p}}_{f,l})-
\eta_{r,0}(\bbox{p}_{f,r})}{\displaystyle 2}\right)
\enspace .
\label{r}
\end{equation}
This boundary condition holds for any value of the transmission
coefficient. In the tunneling limit, $D\ll 1$, it is convenient
to transform Eq.(\ref{r}) to the following equivalent relation
$$
D\sin\left(\frac{\displaystyle \eta_{l,0}(\bbox{p}_{f,l})-
\eta_{r,0}(\bbox{p}_{f,r})}{\displaystyle 2}\right)
\sin\left(\frac{\displaystyle \eta_{l,0}(\underline{\bbox{p}}_{f,l})-
\eta_{r,0}(\underline{\bbox{p}}_{f,r})}{\displaystyle 2}\right)=
\qquad\qquad\qquad\qquad\qquad\qquad
$$
\begin{equation}
\qquad\qquad\qquad\qquad\qquad\qquad\qquad
\sin\left(\frac{\displaystyle \eta_{l,0}(\bbox{p}_{f,l})-
\eta_{l,0}(\underline{\bbox{p}}_{f,l})}{\displaystyle 2}\right)
\sin\left(\frac{\displaystyle \eta_{r,0}(\bbox{p}_{f,r})-
\eta_{r,0}(\underline{\bbox{p}}_{f,r})}{\displaystyle 2}\right)
\enspace .
\label{t}
\end{equation}
It is remarkable that the boundary condition for the quantity
$\eta$ can be formulated separately from other quantities, simply as a
boundary condition for equation (\ref{pe}). Other boundary relations for
quantities entering ansatz (\ref{symmetry}) are given in Appendix.

Relation (\ref{r}) can also be derived within a more general framework
independent of whether a descrete or continuous spectrum is considered.
For this purpose one can represent the Andreev amplitudes in the
form
\begin{equation}
\left(\begin{array}{c} u(\bbox{p}_f,x)\\ v(\bbox{p}_f,x)\end{array}
\right)=\left(\begin{array}{c}
e^{i\eta({\bf p}_f,x)/2}\\ e^{-i\eta({\bf p}_f,x)/2}\end{array}\right)
e^{i\xi({\bf p}_f,x)}
\enspace ,
\label{and}
\end{equation}
where $\eta(\bbox{p}_f,x)$ and $\xi(\bbox{p}_f,x)$ are, in general, complex.
Substituting (\ref{and}) into the boundary conditions for Andreev
amplitudes\cite{shel}, one obtains Eq.(\ref{r}) as a separate boundary
condition for $\eta(\bbox{p}_f,x)$\cite{ira}. In the particular case of
descrete
spectrum $\eta(\bbox{p}_f,x)$ is a real quantity, while $\xi(\bbox{p}_f,x)$ is
purely imaginary leading to exponentially decaying asymptotic behavior of the
Andreev amplitudes. Furthermore, functions entering the expression for the
quasiclassical matrix Green function and satisfying the Riccati equation can,
in general, be represented as
$u(\bbox{p}_f,x)/v(\bbox{p}_f,x)=e^{i\eta(\bbox{p}_f,x)}$\cite{nag,schop}.
Boundary conditions for these functions are evidently directly related with
Eq. (\ref{r}). It is worth noting, in addition, that Eq.(\ref{pe}) transforms
to the Riccati form by introducing the new function $\beta=\tan(\eta/2)$.

If $\eta(\bbox{p}_f,x)$ is a solution of Eqs.\ (\ref{pe}), (\ref{t}) with
the energy $\varepsilon(\bbox{p}_f)$, then $\pi-\eta(\bbox{p}_f,x)$ is a
solution of the same equations with $-\varepsilon(\bbox{p}_f)$ for the
system with a given $\bbox{p}_{f}$ and complex
conjugated order parameter. So, for a given $\bbox{p}_{f}$
quasiparticle descrete spectrum of a system can be, generally speaking,
asymmetric with respect to the Fermi surface under the condition $\phi\ne 0$.

\subsection{Positions of poles and residue values}

As it is shown in this section, energies of bound states, entering denominators
of pole-like terms in the expressions for the propagators, on the one hand, and
residues of those pole-like terms, on the other hand, can be expressed via
$\eta(\bbox{p}_f,x)$ for a given $\Delta(\bbox{p}_f,x)$.

Since $\partial_x\eta(\bbox{p}_f,x)$ vanishes in the bulk
($x\rightarrow\pm\infty$), one immediately gets from Eqs.\ (\ref{pe}) the
relation between the bound state energy $\varepsilon_B(\bbox{p}_f)$ and the
quantities $\eta_{\infty}(\bbox{p}_{f})$, $\Delta_\infty (\bbox{p}_f)$ in
the bulk of the superconductor:
$$
\varepsilon_B(\bbox{p}_{f,r})=|\Delta^r_\infty(\bbox{p}_{f,r})|
\cos\left(\eta_{r,\infty}(\bbox{p}_{f,r})-
\phi_{r,\infty}(\bbox{p}_{f,r})\right)
=\!|\Delta^r_\infty(\underline{\bbox{p}}_{f,r})|
\cos\left(\eta_{r,\infty}(\underline{\bbox{p}}_{f,r})
-\phi_{r,\infty}(\underline{\bbox{p}}_{f,r})\right)
$$
\begin{equation}
\quad\qquad\quad
=|\Delta^l_\infty(\bbox{p}_{f,l})|
\cos\left(\eta_{l,\infty}(\bbox{p}_{f,l})
-\phi_{l,\infty}(\bbox{p}_{f,l})
\right)
=|\Delta^l_\infty(\underline{\bbox{p}}_{f,l})|
\cos\left(\eta_{l,\infty}(\underline{\bbox{p}}_{f,l})
-\phi_{l,\infty}(\underline{\bbox{p}}_{f,l})\right)
.
\label{hphi}
\end{equation}
As a consequence, bound states might exist for a given momentum direction only
below the band edges for the momenta $\bbox{p}_{f,l(r)}$ and
$\underline{\bbox{p}}_{f,l(r)}$, i.e., for
\begin{equation}
|\varepsilon_B(\bbox{p}_f)|\leq
{\rm min}\ \left\{|\Delta^l_\infty(\bbox{p}_{f,l})|,
|\Delta^r_\infty(\bbox{p}_{f,r})|,
|\Delta^l_\infty(\underline{\bbox{p}}_{f,l})|,
|\Delta^r_\infty(\underline{\bbox{p}}_{f,r})|\right\} \enspace .
\label{reg}
\end{equation}

Furthermore, for a frequency near the bound state energy
$\varepsilon_B(\bbox{p}_f)$, the quasiclassical Green's functions can be
expanded in powers of $(\omega-\varepsilon_B(\bbox{p}_f))$.  Taking into
account ansatz (\ref{symmetry}), one has
\begin{equation}
g(\bbox{p}_f,x;\omega)=\frac{\displaystyle
\tilde g(\bbox{p}_f,x;\varepsilon_B(\bbox{p}_f))}{\displaystyle
(\omega-\varepsilon_B(\bbox{p}_f)+i\delta)}+
\sum\limits_{n=0}g^{(n)}(\bbox{p}_f,x;\varepsilon_B(\bbox{p}_f))
(\omega-\varepsilon_B(\bbox{p}_f))^n \enspace ,
\label{g}
\end{equation}
\begin{equation}
f(\bbox{p}_f,x;\omega)=-\frac{\displaystyle
\tilde g(\bbox{p}_f,x;\varepsilon_B(\bbox{p}_f)) }{\displaystyle
(\omega-\varepsilon_B(\bbox{p}_f)+i\delta)}e^{i\eta(
\bbox{p}_f,x)}+
\sum\limits_{n=0}f^{(n)}(\bbox{p}_f,x;\varepsilon_B(\bbox{p}_f))
(\omega-\varepsilon_B(\bbox{p}_f))^n \enspace ,
\label{f_}
\end{equation}
\begin{equation}
f^+(\bbox{p}_f,x;\omega)=\frac{\displaystyle
\tilde g(\bbox{p}_f,x;\varepsilon_B(\bbox{p}_f)) }{\displaystyle
(\omega-\varepsilon_B(\bbox{p}_f)+i\delta)}e^{-i\eta(\bbox{p}_f,x)}+
\sum\limits_{n=0}f^{+ \,(n)}(\bbox{p}_f,x;\varepsilon_B(\bbox{p}_f))
(\omega-\varepsilon_B(\bbox{p}_f))^n \enspace .
\label{ff}
\end{equation}

Further I introduce the quantities
\begin{equation}
f_\pm(\bbox{p}_f,x;\omega)=\frac{\displaystyle 1}{\displaystyle 2}\left(
f^+(\bbox{p}_f,x;\omega)e^{i\eta(\bbox{p}_f,x)}\pm f(\bbox{p}_f,x;\omega)
e^{-i\eta(\bbox{p}_f,x)}\right) \enspace .
\label{f1}
\end{equation}

According to (\ref{g})-(\ref{ff}), $f_+(\bbox{p}_f,x;\omega)$ has no singular
part (no pole-like term), while the singular part of $f_-(\bbox{p}_f,x;\omega)$
coincides with the one for $g(\bbox{p}_f,x;\omega)$, that is
\begin{equation}
\tilde f_-(\bbox{p}_f,x;\varepsilon_B(\bbox{p}_f)) =
\tilde g(\bbox{p}_f,x;\varepsilon_B(\bbox{p}_f)) \enspace , \qquad
\tilde f_+(\bbox{p}_f,x;\varepsilon_B(\bbox{p}_f))=0
\enspace .
\label{f2s}
\end{equation}
In addition, substituting  expansions (\ref{g})-(\ref{ff}) into the
normalization condition (\ref{norm}) and equating terms inversly proportional
to $(\omega-\varepsilon_B(\bbox{p}_f))$, one finds
\begin{equation}
f_{-}^{(0)}(\bbox{p}_f,x;\varepsilon_B(\bbox{p}_f))=
g^{(0)}(\bbox{p}_f,x;\varepsilon_B(\bbox{p}_f)) \enspace .
\label{f2r}
\end{equation}

Taking into account (\ref{f1})--(\ref{f2r}) one derives at the following
relationship (for $x>0$) from the Eilenberger equations
\begin{equation}
f_+^{(0)}(\bbox{p}_f,x;\varepsilon_B(\bbox{p}_f))=\frac{\displaystyle 2i}{
\displaystyle v_x}\int\limits_x^\infty dx\tilde g(\bbox{p}_f,x;
\varepsilon_B(\bbox{p}_f))+f_{+,\infty}^{(0)}(\bbox{p}_f;\varepsilon_B(
\bbox{p}_f)) \enspace .
\label{relat}
\end{equation}

The bulk value of function $f_+^{(0)}(\bbox{p}_f,x;\varepsilon_B(\bbox{p}_f))$
can be easily found:
\begin{equation}
f_{+,\infty}^{(0)}(\bbox{p}_f;\varepsilon_B(\bbox{p}_f))=
-\frac{\displaystyle i\pi|\Delta_{\infty}(\bbox{p}_f)|}{\displaystyle
\sqrt{|\Delta_{\infty}(\bbox{p}_f)|^2-\varepsilon_B^2(\bbox{p}_f)}}
\sin\left(\eta_{\infty}(\bbox{p}_{f})
-\phi_{\infty}(\bbox{p}_{f}) \right) \enspace .
\label{bv}
\end{equation}

Substituting Eqs.(\ref{gx}), (\ref{bv}) into (\ref{relat}) I get
$$
f_+^{(0)}(\bbox{p}_f,x=0;\varepsilon_B(\bbox{p}_f))=i\ {\rm sgn}(v_x)\
|\tilde\Delta^{-1}(\bbox{p}_f,0)|\
\tilde g_0(\bbox{p}_f,\varepsilon_B(\bbox{p}_f))-
\qquad\qquad\qquad\qquad
$$
\begin{equation}
\qquad\qquad\qquad\qquad\qquad
\frac{\displaystyle i\pi |\Delta_{\infty}(\bbox{p}_f)|}{\displaystyle
\sqrt{|\Delta_{\infty}(\bbox{p}_f)|^2-\varepsilon_B^2(\bbox{p}_f)}}
\sin\left(\eta_{\infty}(\bbox{p}_{f})
-\phi_{\infty}(\bbox{p}_{f}) \right) \enspace ,
\label{relat2}
\end{equation}
where
$$
\frac{\displaystyle 1}{\displaystyle |\tilde{\Delta}(\bbox{p}_{f},0)|}=
 \qquad\qquad\qquad\qquad\qquad\qquad\qquad\qquad\qquad\qquad
\qquad\qquad\qquad \qquad\qquad\qquad
$$
\begin{equation}
\frac{\displaystyle 2}{\displaystyle |{\rm v}_{f,x}(\bbox{p}_{f})|}
\int\limits_0^{\infty}
\exp\left(-\displaystyle\frac{\displaystyle 2}{\displaystyle {\rm
v}_{f,x}
(\bbox{p}_{f})}\int\limits_0^{x_1}
|\Delta(\bbox{p}_{f},
x_2)|\sin\left(\eta(\bbox{p}_{f},x_2)
-\phi(\bbox{p}_{f},x_2)\right)dx_2 \right)dx_1 \enspace .
\label{|effdel|}
\end{equation}

In terms of Eq.(\ref{relat2}), (\ref{|effdel|}), one can easily obtain
expression for the residue of the propagator taken at the impenetrable wall.
According to the boundary conditions {\it for an impenetrable wall}, quantities
$f_+^{(0)}(\bbox{p}_f,x=0; \varepsilon_B(\bbox{p}_f))$ and
$\tilde g(\bbox{p}_f,0;\varepsilon_B(\bbox{p}_f))\equiv \tilde g_0(\bbox{p}_f,
\varepsilon_B(\bbox{p}_f))$, taken for incoming momenta are equal to the same
quantities of the outgoing ones. Then one obtains from (\ref{relat2}),
(\ref{hphi}) and (\ref{asr}) the following expression for the residue $\tilde
g_0(\bbox{p}_f,\varepsilon_B(\bbox{p}_f))$ of the propagator
$g(\bbox{p}_f,x>0;\omega)$ taken at the wall (x=0) and at the bound state
energy:
\begin{equation}
\tilde g_0(\bbox{p}_f,\varepsilon_B(\bbox{p}_f))
=\frac{\displaystyle2\pi|\tilde{\Delta}(\bbox{p}_{f},0)|
|\tilde{\Delta}(\underline{\bbox{p}}_{f},0)|}
{|\tilde{\Delta}(\underline{\bbox{p}}_{f},0)|+
|\tilde{\Delta}(\bbox{p}_{f},0)|}
\enspace .
\label{res}
\end{equation}

The positive sign of the residue stipulates a positive contribution from
each bound state to the angle resolved local density of states at the
wall. In accordance with Eqs. (\ref{res}), (\ref{|effdel|}), the residue is
fully determined by the quantities $\Delta(\bbox{p}_{f},x)$,
$\eta(\bbox{p}_{f},x)$.  The quantity $\eta(\bbox{p}_{f},x)$ obeys differential
equation (\ref{pe}), while the position dependent order parameter
$\Delta(\bbox{p}_{f},x)$ needs to be determined self-consistently after
specifying a particular form of the pairing potential. The problem of
self-consistent spatial dependence of $d$-wave order parameter near the
interface was explicitly studied numerically, for example,
in\cite{buch95,ann98,fog98}. The self-consistent space dependent order
parameter $\Delta(\bbox{p}_{f},x)$ is considered to be given throughout this
article.

For the particular case of midgap states and real order parameter one
has\cite{bbs}\  $\varepsilon_B(\bbox{p}_f)=0$, ${\rm sgn
}[\Delta_\infty(\bbox{p}_f)]=-{\rm sgn
}[\Delta_\infty(\underline{\bbox{p}}_f)]$,
$\eta(\bbox{p}_{f},x)=\eta_\infty(\bbox{p}_{f})=
\frac{\displaystyle\pi}{\displaystyle2}{\rm sgn}\left[{\rm
v}_{f,x}(\bbox{p}_{f})\right]+ \phi_\infty(\bbox{p}_{f})$,
$\phi(\bbox{p}_{f},x)=\phi_\infty(\bbox{p}_{f})=
\frac{\displaystyle\pi}{\displaystyle2}\left(1-{\rm sgn
}[\Delta_\infty(\bbox{p}_f)]\right)$. Expressions (\ref{|effdel|}),
(\ref{res}) then reduce to those obtained earlier in Ref.\ \onlinecite{bs}.

\section{Bound states in SIS' junctions}

In order to calculate a bound state energy within the framework of the approach
developed in the preceding section, one should first find
solutions to equation (\ref{pe}), which satisfy both asymptotic relations
(\ref{asr}), (\ref{asl}) and the boundary condition (\ref{t}). The bound state
energy $\varepsilon_B(\bbox{p}_f)$ is directly associated with the asymptotic
value $\eta_\infty(\bbox{p}_{f})$ according to Eq.(\ref{hphi}).

Explicit analytical expressions for the bound state energies
$\varepsilon_B(\bbox{p}_f)$ will be found below under certain conditions in
terms of spatially dependent order parameters on both sides of the interface.
As was demonstrated in Ref.\ \onlinecite{bbs} for the particular case of a
massive supeconductor confined by impenetrable wall, effects of surface pair
breaking for various (although not for all) quasiparticle trajectories result
only in weak deviations of the space dependent quantity $\eta(\bbox{p}_{f},x)$
from its asymptotic value $\eta_{\infty}(\bbox{p}_{f})$:
$|\delta\eta(\bbox{p}_{f},x)|=|\eta(\bbox{p}_{f},x)-
\eta_{\infty}(\bbox{p}_{f})|\ll 1$. In the case of finite transparency of a
junction of two halfspaces one can assume this condition as well, linearize
equation (\ref{pe}) with respect to $\delta\eta(\bbox{p}_{f},x)$ and find the
solutions of Eq.(\ref{pe}) on account of asymptotic conditions (\ref{asr}),
(\ref{asl}) and Eq.(\ref{hphi}):
$$
\eta_l(\bbox{p}_{f,l},x)=\phi_l(\bbox{p}_{f,l})
-{\rm sgn} ({\rm v}^l_{f,x}(\bbox{p}_{f,l}))
\arccos\left(\frac{\displaystyle \varepsilon_B(\bbox{p}_{f,l})}{\displaystyle
|\Delta^l_\infty(\bbox{p}_{f,l})|}\right)+
\frac{\displaystyle 2\varepsilon_B(\bbox{p}_{f,l})}{\displaystyle
{\rm v}^l_{f,x}(\bbox{p}_{f,l})}
\int^x_{-\infty}dx_1
\times
$$
\begin{equation}
\left(1-\frac{\displaystyle |\Delta^l(\bbox{p}_{f,l},x_1)|}{
\displaystyle |\Delta^l_\infty(\bbox{p}_{f,l})|}\right)
\exp\left(-\frac{\displaystyle 2}{\displaystyle
|{\rm v}^l_{f,x}(\bbox{p}_{f,l})|}
\sqrt{1-\frac{\varepsilon^2_B(\bbox{p}_{f,l})
}{|\Delta^l_\infty(\bbox{p}_{f,l})|^2}}
\int\limits^x_{x_1}|\Delta^l(\bbox{p}_{f,l}, x_2)|dx_2\right)
\enspace ,
\label{sol1l}
\end{equation}

$$
\eta_r(\bbox{p}_{f,r},x)=\phi_r(\bbox{p}_{f,r})+
{\rm sgn} ({\rm v}^r_{f,x}(\bbox{p}_{f,r}))
\arccos\left(\frac{\displaystyle \varepsilon_B(\bbox{p}_{f,l})}{\displaystyle
|\Delta^r_\infty(\bbox{p}_{f,r})|}\right)-
\frac{\displaystyle 2\varepsilon_B(\bbox{p}_{f,l})}{\displaystyle
{\rm v}^r_{f,x}(\bbox{p}_{f,r})}
\int_x^{+\infty}dx_1
\times
$$
\begin{equation}
\left(1-\frac{\displaystyle |\Delta^r(\bbox{p}_{f,r},x_1)|}{
\displaystyle |\Delta^r_\infty(\bbox{p}_{f,r})|}\right)
\exp\left(-\frac{\displaystyle 2}{\displaystyle
|{\rm v}^r_{f,x}(\bbox{p}_{f,r})|}
\sqrt{1-\frac{\varepsilon^2_B(\bbox{p}_{f,l})
}{|\Delta^r_\infty(\bbox{p}_{f,r})|^2}}
\int\limits_x^{x_1}|\Delta^r(\bbox{p}_{f,r}, x_2)|dx_2\right)
\enspace .
\label{sol1r}
\end{equation}

Effects of supercurrent flowing across the junction (along the $x$ axis) can be
taken into account by adding the spatially depending term $2mv_sx$ into the
phases $\phi_{l(r)}(\bbox{p}_{f,l(r)})$ in (\ref{pe}). Then the corresponding
solutions are obtained from (\ref{sol1l}), (\ref{sol1r}) after the substitution
$\varepsilon_B(\bbox{p}_{f,l})\to \varepsilon_B(\bbox{p}_{f,l})
- {\bf v}_{f}^{l(r)}(\bbox{p}_{f,l(r)})m\bbox {v}_{s}$,\
$\phi_{l(r)}(\bbox{p}_{f,l(r)})\to \phi_{l(r)}(\bbox{p}_{f,l(r)})+2mv_sx$.

Bound state energies can be found now combining solutions (\ref{sol1l}),
(\ref{sol1r}) and the boundary condition (\ref{t}). The equation
obtained within the approach should be linearized with respect to the effects
of surface pair breaking ( discribed by spatial integrals in (\ref{sol1l}),
(\ref{sol1r})) both in the right and the left superconductors.

\subsection{Low energy bound states}

Let us consider bound states with low energies
$$\varepsilon_B(\bbox{p}_{f,l})\ll{\rm min}\left\{
|\Delta^l_\infty(\bbox{p}_{f,l})|, |\Delta^l_\infty(
\underline{\bbox{p}}_{f,l})|, |\Delta^r_\infty(\bbox{p}_{f,r})|,
|\Delta^r_\infty(\underline{\bbox{p}}_{f,r})|\right\} \enspace .$$

Linearizing Eqs.(\ref{sol1l}), (\ref{sol1r}) with respect to the small
parameters
$\varepsilon_B(\bbox{p}_{f,l})/|\Delta^{l(r)}_\infty(\bbox{p}_{f,l(r)})|$ and
taking into account the presence of a supercurrent, one gets after simple
transformations:
\begin{eqnarray}
\eta_l(\bbox{p}_{f,l},x)=\phi_l(\bbox{p}_{f,l})
-\frac{\displaystyle \pi}{\displaystyle 2}
{\rm sgn} ({\rm v}^l_{f,x}(\bbox{p}_{f,l}))+
\frac{\displaystyle \left(\varepsilon_B(\bbox{p}_{f,l})
- {\bf v}_{f}^{l}(\bbox{p}_{f,l})m\bbox {v}_{s}\right)
{\rm sgn} ({\rm v}^l_{f,x}(\bbox{p}_{f,l}))}{\displaystyle
|\tilde{\Delta}^l(\bbox{p}_{f,l},x)|} \enspace ,\label{linsol}\\
\eta_r(\bbox{p}_{f,r},x)=\phi_r(\bbox{p}_{f,r})
+\frac{\displaystyle \pi}{\displaystyle 2}
{\rm sgn} ({\rm v}^r_{f,x}(\bbox{p}_{f,r}))-
\frac{\displaystyle \left(\varepsilon_B(\bbox{p}_{f,l})
- {\bf v}_{f}^{r}(\bbox{p}_{f,r})m\bbox {v}_{s}\right)
{\rm sgn} ({\rm v}^r_{f,x}(\bbox{p}_{f,r}))}{\displaystyle
|\tilde{\Delta}^r(\bbox{p}_{f,r},x)|} \enspace ,
\label{linsor}
\end{eqnarray}
where the following quantities are introduced
\begin{eqnarray}
\frac{\displaystyle 1}{\displaystyle |\tilde{\Delta}^l(\bbox{p}_{f},x)|}=
\frac{\displaystyle 2}{\displaystyle |{\rm v}^l_{f,x}(\bbox{p}_{f})|}
\int^x_{-\infty}
\exp\left(-\displaystyle\frac{\displaystyle 2}{\displaystyle |{\rm v}^l_{f,x}
(\bbox{p}_{f})|}\int\limits^x_{x_1}
|\Delta^l(\bbox{p}_{f}, x_2)|dx_2\right)dx_1 \enspace ,  \label{effdel}\\
\frac{\displaystyle 1}{\displaystyle |\tilde{\Delta}^r(\bbox{p}_{f},x)|}=
\frac{\displaystyle 2}{\displaystyle |{\rm v}^r_{f,x}(\bbox{p}_{f})|}
\int_x^{\infty}
\exp\left(-\displaystyle\frac{\displaystyle 2}{\displaystyle |{\rm v}^r_{f,x}
(\bbox{p}_{f})|}\int\limits_x^{x_1}
|\Delta^r(\bbox{p}_{f}, x_2)|dx_2\right)dx_1 \enspace .
\label{effder}
\end{eqnarray}

I first assume that {\it in the limit $D\rightarrow 0$ midgap states arise on
both sides of the barrier plane} for given crystal to surface orientations and
quasiparticle trajectories considered. For simplicity, let the phases of the
order parameters $\phi_l(\bbox{p}_{f,l})$, $\phi_r(\bbox{p}_{f,r})$ in both
superconducors be spatially constant in the absence of a supercurrent.

Then, allowing for the difference $\phi$  between the phases of right and left
comlex order parameters, these phases may be written under the conditions
considered as follows
\begin{equation}
\begin{array}{ll}
\phi_l(\bbox{p}_{f,l})=\phi_l+2mv_sx \enspace ,&
\phi_r(\bbox{p}_{f,r})=\phi_r+
\pi\delta_{i1}
+2mv_sx \enspace , \\
\phi_l(\underline{\bbox{p}}_{f,l})=\phi_l+
\pi+2mv_sx
\enspace ,&
\phi_r(\underline{\bbox{p}}_{f,r})=\phi_r-
\pi\delta_{i2}
+2mv_sx
\enspace ,
\end{array}
\label{chi}
\end{equation}
where two different cases $i=1,2$ are taken into account. The superfluid
velocity $v_s$ is assumed to be positive (negative) for the supercurrent
flowing along (opposite to) the $x$-axis.

Expanding boundary condition (\ref{t}) with the substitution (\ref{linsol}),
(\ref{linsor}) with respect to $\left(\varepsilon_B(\bbox{p}_{f,l}) - m\bbox
{v}_{s}{\bf v}_{f}^{l(r)}(\bbox{p}_{f,l(r)})\right)$, results in a simple
quadratic equation for the bound state energies:
$$
\left[\varepsilon_B(\bbox{p}_{f,r})
\frac{\displaystyle |\tilde{\Delta}^l(\bbox{p}_{f,l},0)|+|\tilde{\Delta}^l(
\underline{\bbox{p}}_{f,l},0)|}{\displaystyle
|\tilde{\Delta}^l(\bbox{p}_{f,l},0)||\tilde{\Delta}^l(
\underline{\bbox{p}}_{f,l},0)|}{\rm sgn}({\rm v}^l_{f,x}(\bbox{p}_{f,l}))
-mv_s\left(\frac{\displaystyle |{\rm v}^l_{f,x}(\bbox{p}_{f,l})|
}{\displaystyle |\tilde{\Delta}^l(\bbox{p}_{f,l},0)|}-
\frac{\displaystyle |{\rm v}^l_{f,x}(\underline{\bbox{p}}_{f,l})|
}{\displaystyle |\tilde{\Delta}^l(
\underline{\bbox{p}}_{f,l},0)|}\right)
\right]\times
$$
$$
\left[\varepsilon_B(\bbox{p}_{f,r})
\frac{\displaystyle |\tilde{\Delta}^r(\bbox{p}_{f,r},0)|+|\tilde{\Delta}^r(
\underline{\bbox{p}}_{f,r},0)|}{\displaystyle
|\tilde{\Delta}^r(\bbox{p}_{f,r},0)||\tilde{\Delta}^r(
\underline{\bbox{p}}_{f,r},0)|}{\rm sgn}({\rm v}^r_{f,x}(
\underline{\bbox{p}}_{f,r}))
+mv_s\left(\frac{\displaystyle |{\rm v}^r_{f,x}(\bbox{p}_{f,r})|
}{\displaystyle |\tilde{\Delta}^r(\bbox{p}_{f,r},0)|}-
\frac{\displaystyle |{\rm v}^r_{f,x}(\underline{\bbox{p}}_{f,r})|
}{\displaystyle |\tilde{\Delta}^r(
\underline{\bbox{p}}_{f,r},0)|}\right)
\right]
$$
\begin{equation}
\qquad\qquad\qquad\qquad\qquad\qquad\qquad\qquad
=4D\sin\left(\frac{\displaystyle \phi}{\displaystyle 2}+
\frac{\displaystyle \pi}{\displaystyle 2}\delta_{i1}
\right)
\cos\left(\frac{\displaystyle \phi}{\displaystyle 2}-
\frac{\displaystyle \pi}{\displaystyle 2}\delta_{i2}
\right)
\enspace .
\label{quadr}
\end{equation}
For the sake of simplicity let $D\ll1$ in Eq.(\ref{quadr}).

Two particular solutions of Eq.(\ref{quadr}) are of special interest.
For a {\it ``symmetric'' tunnel junction} (STJ) identical superconductors with
the same orientations are situated on its left and right sides.
Subscript $i=1$ in Eqs.(\ref{chi}), (\ref{quadr}) corresponds to this case.
Absolute values of order parameter on both sides of the STJ
taken at the same distance from the interface, are equal to each other, even
in the presence of a phase difference and supercurrent:
$|\Delta^l(\bbox{p}_{f,l},-x)|=|\Delta^r(\underline{\bbox{p}}_{f,r},x)|$,
$|\Delta^l(\underline{\bbox{p}}_{f,l},-x)|=|\Delta^r(\bbox{p}_{f,r},x)|$.
Then analogous equalities follow from Eqs.(\ref{effdel}), (\ref{effder})
for effective order parameters:
$|\tilde\Delta^l(\bbox{p}_{f,l},-x)|=|\tilde\Delta^r(\underline{\bbox{p}}_{f,r}
,x)|$,
$|\tilde\Delta^l(\underline{\bbox{p}}_{f,l},-x)|=|\tilde\Delta^r(\bbox{p}_{f,r}
,x)|$.
Under these conditions one gets
$$
\varepsilon_B^{STJ}(\bbox{p}_{f,r})=
\pm\frac{\displaystyle 2|\tilde\Delta^r(\bbox{p}_{f,r},0)||\tilde\Delta^r(
\underline{\bbox{p}}_{f,r},0)|}{\displaystyle |\tilde\Delta^r(\bbox{p}_{f,r},0)
|+|\tilde\Delta^r(\underline{\bbox{p}}_{f,r},0)|}\sqrt{\displaystyle D}
\left|\cos\frac{\displaystyle \phi}{\displaystyle 2}\right|+
\qquad\qquad\qquad\qquad\qquad\qquad\qquad\qquad
$$
\begin{equation}
\qquad\qquad\qquad\qquad\qquad\qquad\qquad
\frac{\displaystyle {\rm v}^r_{f,x}(\bbox{p}_{f,r})|\tilde\Delta^r(\underline{
\bbox{p}}_{f,r},0)|+{\rm v}^r_{f,x}(\underline{\bbox{p}}_{f,r})|\tilde\Delta^r(
\bbox{p}_{f,r},0)|}{
\displaystyle |\tilde\Delta^r(\bbox{p}_{f,r},0)| +|\tilde\Delta^r(\underline{
\bbox{p}}_{f,r},0)|}mv_s \enspace .
\label{estj}
\end{equation}

For a {\it ``mirror'' tunnel junction} (MTJ) $i=2$ orientations of
identical superconductors can be obtained from each other, by definition, by
a reflection with respect to the junction barrier plane. Then
$|\Delta^l(\bbox{p}_{f,l},-x)|=|\Delta^r(\bbox{p}_{f,r},x)|$,
 $|\Delta^l(\underline{\bbox{p}}_{f,l},-x)|=|\Delta^r(
\underline{\bbox{p}}_{f,r},x)|$ and one obtains the corresponding values of
bound state energy from Eqs.(\ref{effdel}), (\ref{effder}), (\ref{quadr}):
$$
\varepsilon_B^{MTJ}(\bbox{p}_{f,r})=
\qquad\qquad\qquad\qquad\qquad\qquad\qquad\qquad\qquad\qquad\qquad\qquad\qquad
\qquad\qquad\qquad\qquad
$$
\begin{equation}
\pm\frac{\displaystyle |\tilde\Delta^r(\bbox{p}_{f,r},0)||\tilde\Delta^r(
\underline{\bbox{p}}_{f,r},0)|}{\displaystyle |\tilde\Delta^r(\bbox{p}_{f,r},0)
|+|\tilde\Delta^r(\underline{\bbox{p}}_{f,r},0)|}\sqrt{\displaystyle
4D\sin^2\left(\frac{\displaystyle \phi}{\displaystyle 2}\right)+
\left(\frac{\displaystyle |{\rm v}^r_{f,x}(\bbox{p}_{f,r})|}{
\displaystyle |\tilde\Delta^r(\bbox{p}_{f,r},0)|}-
\frac{\displaystyle |{\rm v}^r_{f,x}(\underline{\bbox{p}}_{f,r})|}{
\displaystyle |\tilde\Delta^r(\underline{\bbox{p}}_{f,r},0)|}\right)^2
(mv_s)^2 } \enspace .
\label{emtj}
\end{equation}

As $v_s$ can depend upon $\varepsilon_B$, relations (\ref{estj}), (\ref{emtj})
still represent, generally speaking, implicit equations for $\varepsilon_B$.
However, factors in front of $mv_s$ in Eqs.(\ref{estj}), (\ref{emtj})  can be
extremely small or even vanish (as in the case
$|\tilde\Delta^r(\underline{\bbox{p}}_{f,r},0)|=|\tilde\Delta^r(\bbox{p}_{f,r},
0)|$, ${\rm v}^r_{f,x}(\underline{\bbox{p}}_{f,r})= -{\rm
v}^r_{f,x}(\bbox{p}_{f,r})$). Then one can disregard the influence of the
superflow on the bound state energy, and get fairly simple results from
Eqs.(\ref{estj}), (\ref{emtj})
\begin{equation}
\varepsilon_B(\bbox{p}_{f,r})=
\pm\frac{\displaystyle 2|\tilde\Delta^r(\bbox{p}_{f,r},0)||\tilde\Delta^r(
\underline{\bbox{p}}_{f,r},0)|}{\displaystyle |\tilde\Delta^r(\bbox{p}_{f,r},0)
|+|\tilde\Delta^r(\underline{\bbox{p}}_{f,r},0)|}\sqrt{\displaystyle D}
\left\{\begin{array}{l}
\left|\cos\frac{\displaystyle \phi}{\displaystyle 2}\right| \qquad for\ STJ ,\\
\\
\left|\sin\frac{\displaystyle \phi}{\displaystyle 2}\right| \qquad for\ MTJ\ .\\
\end{array}
\right.
\label{levels}
\end{equation}

One should note the relation $\varepsilon_B\propto\sqrt{D}$  and the
essentially different dependences of the bound state energies upon the phase
difference $\phi$ for STJ and MTJ. According to (\ref{estj})-(\ref{levels}),
zero energy bound states at an impenetrable wall ($D\to 0$) in the absence of
supercurrent, split into two low energy levels on account of the effects of
nonzero (low) transmission and phase difference. For STJ the bound state energy
has its maximum for $\phi=0$ and the principal reason for the bound state
energy shifting away from the midgap position is the finite transmission.  For
MTJ the bound state energy differs from zero for nonzero $\phi$ taking its
maximum value at $\phi=\pi$. This difference is evidently related to the
additional phase shift $\pi$ acquired by paired quasiparticles crossing the
junction and seeing different signs of the gap functions for the given momentum
direction in two superconductors in MTJ. Let now midgap states take place {\it
in the limit $D\rightarrow 0$ only on one side of the barrier plane}, for
instance, in the right superconductor for quasiparticle trajectories
considered. Then expressions for the phases of complex order parameters differ
from Eq.(\ref{chi}) and have the following form
\begin{equation}
\begin{array}{ll}
\phi_l(\bbox{p}_{f,l})=\phi_l+2mv_sx \enspace ,&
\phi_r(\bbox{p}_{f,r})=\phi_r+
\pi\delta_{i1}
+2mv_sx \enspace , \\
\phi_l(\underline{\bbox{p}}_{f,l})=\phi_l + 2mv_sx
\enspace ,&
\phi_r(\underline{\bbox{p}}_{f,r})=\phi_r-
\pi\delta_{i2}
+2mv_sx
\enspace .
\end{array}
\label{chi2}
\end{equation}

In the case of an impenetrable wall there is no solution to equation (\ref{pe})
in the left half space $x<0$ for sufficiently small energy. Such a solution
arises, however, at finite transmission, being induced by the proximity effect
on account of the corresponding solution for the right half space. These
solutions have the same form (\ref{sol1l}), (\ref{sol1r}) with new relations
(\ref{chi2}).

Boundary condition (\ref{t}) then reduces to
$$
\left[\eta_{r,0}(\bbox{p}_{f,r})-
\phi_r(\bbox{p}_{f,r})
-\frac{\displaystyle \pi}{\displaystyle 2}
{\rm sgn} ({\rm v}^r_{f,x}(\bbox{p}_{f,r}))\right]-
\left[\eta_{r,0}(\underline{\bbox{p}}_{f,r})
-\phi_r(\underline{\bbox{p}}_{f,r})
-\frac{\displaystyle \pi}{\displaystyle 2}
{\rm sgn} ({\rm v}^r_{f,x}(\underline{\bbox{p}}_{f,r}))\right]
\qquad\quad
$$
\begin{equation}
\qquad\quad\qquad\qquad\qquad\qquad\qquad\qquad\qquad\qquad
\qquad\qquad\qquad\qquad\qquad
=\left(-1\right)^{\displaystyle\delta_{i1}}D\sin\phi
\enspace .
\label{deltatr}
\end{equation}

In terms of Eqs.(\ref{deltatr}), (\ref{ll})-(\ref{rlp}), one can justify that,
indeed, $\tilde{g}_{l,0}\propto D\tilde{g}_{r,0}$ both for the incoming and
outgoing momentum directions.

Substituting $\eta_{r,0}(\bbox{p}_{f,r})$,
$\eta_{r,0}(\underline{\bbox{p}}_{f,r})$ from Eq.(\ref{linsor}) into
Eq.(\ref{deltatr}), one gets the following expression for the bound state
energy
$$
\varepsilon_B(\bbox{p}_{f,r})=
\frac{\displaystyle |\tilde\Delta^r(\bbox{p}_{f,r},0)||\tilde
\Delta^r(\underline{\bbox{p}}_{f,r},0)|{\rm sgn}\left({\rm v}^r_{f,x}(
\underline{\bbox{p}}_{f,r})\right)}{\displaystyle |\tilde\Delta^r(
\bbox{p}_{f,r},0)|+|\tilde\Delta^r(\underline{\bbox{p}}_{f,r},0)|}
\left[ \left(-1\right)^{\displaystyle\delta_{i1}}D\sin\phi-
\right.
\qquad\qquad\qquad\qquad\qquad
$$
\begin{equation}
\left.
\qquad\qquad\qquad\qquad\qquad\qquad\qquad\qquad\qquad\qquad
mv_s\left(\frac{\displaystyle |{\rm v}^r_{f,x}(\bbox{p}_{f,r})|}{
\displaystyle |\tilde\Delta^r(\bbox{p}_{f,r},0)|}-
\frac{\displaystyle |{\rm v}^r_{f,x}(\underline{\bbox{p}}_{f,r})|}{
\displaystyle |\tilde\Delta^r(\underline{\bbox{p}}_{f,r},0)|}\right)\right]
\enspace .
\label{epb}
\end{equation}

Results obtained in this section show that the effect of surface pair breaking
on the low-energy bound states, can be taken into account by introducing
the effective surface values of order parameters definded in Eq.(\ref{effdel}),
(\ref{effder}). If one disregards for the time being the surface pair breaking
at all, considering moduli of the order parameters being independent of spatial
coordinates near interfaces, then it follows from Eqs.(\ref{effdel}),
(\ref{effder}) \,
$|\tilde{\Delta}(\bbox{p}_{f},x)|=|\Delta_\infty(\bbox{p}_{f})|=const(x)$. For
crystalline orientations, when $|\Delta_\infty(\bbox{p}_{f,r})|=
|\Delta_\infty(\underline{\bbox{p}}_{f,r})|$, one gets from Eq.(\ref{levels})
in this particular case $\varepsilon_B^{STJ}(\bbox{p}_{f,r})=\pm|\Delta_\infty(
\bbox{p}_{f,r})|\sqrt{\displaystyle D}\left|\cos\frac{\displaystyle\phi}{
\displaystyle 2}\right|$, \ $\varepsilon_B^{MTJ}(\bbox{p}_{f,r})=
\pm|\Delta_\infty(\bbox{p}_{f,r})|\sqrt{\displaystyle D}\left|\sin\frac{
\displaystyle\phi}{\displaystyle 2}\right|$.  According to (\ref{effdel}),
(\ref{effder}), the relation
$|\tilde{\Delta}(\bbox{p}_{f},x)|\le|\Delta_\infty(\bbox{p}_{f})|$ holds as a
consequence of surface pair breaking. Disregarding this effect leads to an
overestimation of the shift of the zero energy bound states brought about by
the finite transmission of the junction barrier. Such an overestimation can be
quite noticeable. Moreover, for smooth surfaces the suppression of the order
parameters depends on crystal to surface orientation. The effective surface
quantities (\ref{effdel}), (\ref{effder}) can manifest qualitatively different
dependences upon crystal to interface orientation as compared to the
dependences of the order parameters in the bulk.

\subsection{Splitting from surface pair breaking of the bound states near the
edge of the continuous spectrum}

In junctions of isotropic $s$-wave superconductors, the order parameters with
incoming and reflected quasiparticle momenta are identical. In the $s$-wave
case one can usually disregard surface pair breaking and consider the order
parameter as spatially constant up to the interface. Then the energies of the
interface bound states in symmetric junctions, as it is well known, are as
follows\cite{fur90,been291}:
\begin{equation}
\varepsilon_B(\theta)=\pm|\Delta|\sqrt{1-D(\theta)\sin^2(\phi/2)}
\enspace .
\label{iss}
\end{equation}
It is explicitly indicated here that transmission depends upon the angle
$\theta$ between momentum direction and the interface normal. For tunnel
junctions ($D\ll 1$) energy (\ref{iss}) lies close to the edge of continuous
spectrum.

In preceding section the presence of quasiparticle trajectories was assumed,
where incoming and outgoing quasiparticles see the order parameter values of
opposite signs. The corresponding states have quite low energies in the
tunneling limit. Interface bound states in the vicinity of continuous spectrum,
however, always arise in tunnel junctions between $d$-wave superconductors as
well. This is the case for quasiparticle trajectories with momentum directions
along the interface normal, where incoming and reflected quasiparticles always
see the same order parameter of an anisotropic singlet superconductor.
Interface bound states in $d$-wave superconductors for trajectories in close
vicinity of the interface normal are to some extent analogous to the ones in
isotropic $s$-wave case. For special crystalline orientations such as
$45^\circ$ in $d_{x^2-y^2}$ superconductors all those trajectories
reduce to the only one along the interface normal.

If one considers a contribution of bound states to the current across the
junction, then the particular dependence of a transmission upon $\theta$ turns
out to be especially important. For the transmission gradually diminishing with
increasing $\theta$, like in case of sufficiently thin barriers, trajectories
in a wide range of $\theta$ can substantially contribute to the current. By
contrast, if the transmission noticeably differs from zero only for small
$\theta$ (this is the case for thick barriers), the contribution from
trajectories with sufficiently small $\theta$ entirely governs the current.
Below bound states in junctions between $d$-wave superconductors are
considered just for the trajectory along the interface normal.

Bound states (\ref{iss}) are formed due to Andreev reflection processes
resulting from a phase difference in the order parameters of a symmetric
junction. According to (\ref{iss}), the bound states merge into the continuum
in the limit $D\to 0$. However, this is not actually the case as a result
of a suppression of the moduli of the order parameters in the vicinity of the
barrier plane. For a smooth impenetrable wall, the bound states still exist
on account of surface pair breaking for momentum direction along (or
sufficiently close to) the surface normal for almost all crystal to surface
orientations of a $d$-wave superconductor\cite{bbs}. Bound states of this kind
arise even for a weak suppression of the order parameter. Then, however,
there must be a tiny distance from the energy of a bound state to the
continuum. For isotropic $s$-wave superconductors this applies to any
quasiparticle trajectory.

Inhomogeneous modulus of the order parameter as such near an
impenetrable wall can be considered an effective potential well for
quasiparticles. Finite transmission across the interface of a junction, as
opposed to the impenetrable wall, results in a double
well structure and allows for the interplay of the phase difference and the
inhomogeneous moduli of the order parameters in forming interface bound states. It
is shown below, that splitting of bound state levels takes place in the
double well structure for the quasiparticle trajectory along the surface
normal, at least in the tunneling limit.

One can easily see from (\ref{hphi}) for positive bound state energies near the
continuum, that in ``symmetric'' or ``mirror'' tunnel junctions all asymptotic
values $\left(\eta_{\infty}(\bbox{p}_{f})-\phi_{\infty}(\bbox{p}_{f})
\right)$ are small. Assuming in addition
$|\eta(\bbox{p}_{f},x)-\eta_{\infty}(\bbox{p}_{f})|\ll 1$,
one can get from Eq.(\ref{pe}) in the first approximation the following
expressions for the corresponding interface values $\eta(\bbox{p}_{f},x=0)
\equiv\eta_0(\bbox{p}_{f})$ :
\begin{equation}
\eta_{l,0}(\bbox{p}_{f,l})=-{\rm sgn}({\rm v}^l_{f,x}(
\bbox{p}_{f,l}))\sqrt{1-\frac{\displaystyle \varepsilon^2_B(\bbox{p}_{f,l})}{
\displaystyle |\Delta^l_{-\infty}(\bbox{p}_{f,l})|^2}}
+\frac{\displaystyle 2\varepsilon_B(\bbox{p}_{f,l})}{\displaystyle
{\rm v}^l_{f,x}(\bbox{p}_{f,l})}\int_{-\infty}^0\left(1-\frac{
|\Delta^l(\bbox{p}_{f,l},x)|}{|\Delta^l_{-\infty}(\bbox{p}_{f,l})|}\right)dx
\enspace ,
\end{equation}
\begin{equation}
\eta_{r,0}(\bbox{p}_{f,l})=\phi+{\rm sgn}({\rm v}^r_{f,x}(
\bbox{p}_{f,l}))\sqrt{1-\frac{\displaystyle \varepsilon^2_B(\bbox{p}_{f,l})}{
\displaystyle |\Delta^r_{\infty}(\bbox{p}_{f,r})|^2}}
-\frac{\displaystyle 2\varepsilon_B(\bbox{p}_{f,l})}{\displaystyle
{\rm v}^r_{f,x}(\bbox{p}_{f,r})}\int^{\infty}_0\left(1-\frac{
|\Delta^r(\bbox{p}_{f,r},x)|}{|\Delta^r_{\infty}(\bbox{p}_{f,r})|}\right)dx
\enspace .
\end{equation}

Substituting these expressions into the boundary condition (\ref{t}) and
linearizing in
$\left(\eta_0(\bbox{p}_{f})-\eta_{0}(\underline{\bbox{p}}_{f})\right)$,
one obtains for STJ and MTJ the following bound state
energies:
\begin{equation}
\frac{\displaystyle\varepsilon^2_B(\bbox{p}_{f,r})}{\displaystyle
|\Delta^r_{\infty}(\bbox{p}_{f,r})|^2}=1-\left[
\frac{\displaystyle 2|\Delta^r_{\infty}(\bbox{p}_{f,r})|}{\displaystyle
|{\rm v}^r_{f,x}(\bbox{p}_{f,r})|}\int_0^\infty\left(1-
\frac{\displaystyle |\Delta^r(\bbox{p}_{f,r},x)|}{\displaystyle
|\Delta^r_{\infty}(\bbox{p}_{f,r})|}\right)dx
\pm
\sqrt{\displaystyle D}
\sin\left(\frac{\displaystyle \phi}{\displaystyle2}\right)
\right]^2
\enspace .
\label{split}
\end{equation}
The minus sign in front of the $\sqrt{D}$ - term in (\ref{split}) is admissible
only when the modulus of this term is less than that of the first term in the
square brackets. This implies a relatively large role of interface pair
breaking in forming Andreev bound states as opposed to the contribution from
the phase difference. As can be seen in Eq.(\ref{split}), under these
circumstances the combined effect of the phase difference and the interface
pair breaking results in a splitting of the bound state energy.  In the
opposite limit of absence of noticeable surface pair breaking, when
$|\Delta(\bbox{p}_{f},x)|=|\Delta_{\infty}(\bbox{p}_{f})|$, the first term
within the square brackets in Eq.(\ref{split}) vanishes. Then Eq.(\ref{split})
reduces to the conventional result Eq.(\ref{iss}). On the other hand, for an
impenetrable barrier Eq.(\ref{split}) takes the form found in Ref.\
\onlinecite{bbs}. If order parameters have equal moduli and opposite signs on
the two sides of the junction, then the bound state energies are described by
Eq.(\ref{split}) after replacing $\sin\left(\frac{\displaystyle
\phi}{\displaystyle2}\right)$ by $\cos\left(\frac{\displaystyle
\phi}{\displaystyle2}\right)$.

The splitting considered above is formally analogous to the splitting discussed
in Ref.\ \onlinecite{wen196} for long SIS junctions and specially constructed
SNS junctions with double-well or double-barrier structures. However, the
physical reasons for the splitting and the conditions for its observations
considered in this section, drastically differ from what was studied earlier,
where the
effect of interface pair breaking was always neglected. The double well
structure discussed in this section is inherent in junctions with noticeable
interface pair breaking. One should note that the first term in square brackets
in Eq.(\ref{split}) is considerably less than unity both for $s$-wave
superconductors as well as for $d$-wave superconductors. In the former case
this is entirely due to a small suppression of the order parameter at the
interface. The $d$-wave order parameter taken for a
momentum along the surface normal, is small in itself for those crystal to
interface orientations, which entail substantial surface pair breaking.
However, at least for the $d$-wave case the first term in square brackets
in Eq.(\ref{split})can exceed the second one, allowing an observable splitting
(the estimations follow from Fig.3 in\cite{bbs}).

The supercurrent flowing via individual bound state (\ref{split}) can be
written as
\begin{equation}
I_{\pm}\propto e|\Delta^r_{\infty}(\bbox{p}_{f,r})|\sqrt{D}\cos\left(
\frac{\displaystyle\phi}{\displaystyle2}\right)\left[\sqrt{D}\sin\left(
\frac{\displaystyle\phi}{\displaystyle2}\right)\pm \frac{\displaystyle
2|\Delta^r_{\infty}(\bbox{p}_{f,r})|}{\displaystyle
|{\rm v}^r_{f,x}(\bbox{p}_{f,r})|}\int_0^\infty\left(1-
\frac{\displaystyle |\Delta^r(\bbox{p}_{f,r},x)|}{\displaystyle
|\Delta^r_{\infty}(\bbox{p}_{f,r})|}\right)dx\right]
\enspace .
\end{equation}
For sufficiently small transmission the second term in the square brackets
dominates, resulting in a supercurrent proportional to $\sqrt{D}$.
As it is known for resonance Josephson coupling in double well
structures\cite{wen196}, in equilibrium such anomalous terms coming
collectively from all bound and extended states cancel in the expression for
the total current. For nonequilibrium quasiparticle occupation of bound and
(or) extended states, however, one can expect observable manifestations of the
anomaly discussed above. As compared to the conventional behavior linear in
the transmission coefficient, a $\sqrt{D}$-term results in a inhancement
of nonequilibrium Josephson current from surface pair breaking in the tunneling
limit. For $d$-wave superconductors, the broadening of bound states on
interface roughness and impurities\cite{pbbi} is of importance, since can
smearing out the splitting.

It is not entirely clear whether the $\sqrt{D}$-behavior of the critical
current vanishes in equilibrium within more general assumptions than
made above. Such a study is in progress now.

\section{Broaden bound states in NIS junctions}

Andreev bound states, localized near an impenetrable wall in a d-wave
superconductor, become slightly broadened in the case of small but finite
barrier transmission of an NIS tunnel junction. In this case, the retarded
propagators have no singularities at any real value of the energy , since
positions of poles move from the real axis in the complex energy plane. This
shift is linear in the barrier transmission coefficient, which is assumed to be
sufficiently small. A relatively small imaginary part of a pole position is
associated with finite quasiparticle life time for the quasistationary bound
state.

The other important feature of NIS (normal metal - insulator - superconductor)
junctions is that the solution for the normal metal halfspace cannot be
described by ansantz (\ref{symmetry})\cite{sns,ans}. At the same time the
dominating terms in propagators still satisfy Eq.(\ref{symmetry}) in the
superconducting half space in the case of sufficiently small transparency of
the barrier.

As it follows from Eilenberger's equations, the diagonal component of the
quasiclassical Green's function is constant throughout the normal metal up to
the interface ($g=-i\pi$). At the same time superconducting correlations are
present there due to the proximity effect and known to decrease exponentially
toward the bulk normal metal ($x<0$). Since imaginary part of a pole position
should be negative for the retarded propagator, one can write for the anomalous
Green's functions for $x<0$:
\begin{equation}
\left\{\begin{array}{l}
f_N\left(\bbox{p}_{f,l},x,\varepsilon_B(\bbox{p}_{f,l})\right)
=f_{N0}\left(\bbox{p}_{f,l},\varepsilon_B(\bbox{p}_{f,l})\right)
\exp\left(\displaystyle i\frac{\displaystyle
2\varepsilon_B(\bbox{p}_{f,l})x}{\displaystyle v^l_{f,x}(\bbox{p}_{f,l})}
\right)\ ,
\\
f^+_N\left(\underline{\bbox{p}}_{f,l},x,\varepsilon_B(\bbox{p}_{f,l})\right)
=f^+_{N0}\left(\underline{\bbox{p}}_{f,l},\varepsilon_B(\bbox{p}_{f,l})\right)
\exp\left(\displaystyle -i\frac{\displaystyle2\varepsilon_B(\bbox{p}_{f,l})x}{
\displaystyle v_{f,x}(\underline{\bbox{p}}_{f,l})}\right) \ ,
\end{array}
\right.
\label{f}
\end{equation}
while $
f_N\left(\underline{\bbox{p}}_{f,l},x,\varepsilon_B(\bbox{p}_{f,l})\right)=0$,
\ $f^+_N\left(\bbox{p}_{f,l},x,\varepsilon_B(\bbox{p}_{f,l})\right)=0$\ .
Here and below normal to the interface velocity component
$v^l_{f,x}(\bbox{p}_{f,l})$ \ (~$v^l_{f,x}(\underline{\bbox{p}}_{f,l})$~) is
chosen to be positive (~negative~) for an incoming (~outgoing~) quasiparticle
in the left halfspace.

Evidently, propagators (\ref{f}) in the left halfspace do not satisfy relation
(\ref{symmetry}). There is, however, another important relation, which strictly
holds in the normal metal halfspace and substantially simplificates Zaitsev's
boundary conditions at the NIS interface:  diagonal components of matrix
$\hat{d}_l$ are equal to zero in the case considered. According to
(\ref{zai2}), diagonal components of $\hat{d}_r$ vanish in this case as well.
Thus, applying ansatz (\ref{symmetry}) to the right halfspace and taking into
account (\ref{zai2}), (\ref{f}), one gets
\begin{eqnarray}
{f}_{N0}(\bbox{p}_{f,l},\varepsilon_B(\bbox{p}_{f,l}))=
g_{r,0} (\bbox{ p}_{f,r},\varepsilon_B(\bbox{p}_{f,l}))\left[
e^{\displaystyle i\eta_{r,0}(\bbox{p}_{f,r})}-e^{\displaystyle i
\eta_{r,0}(\underline{\bbox{p}}_{f,r})}\right]
\enspace , \nonumber \\
\nonumber\\
{f}^+_{N0}(\underline{\bbox{p}}_{f,l},\varepsilon_B(\bbox{p}_{f,l}))=
g_{r,0} (\bbox{ p}_{f,r},\varepsilon_B(\bbox{p}_{f,l}))\left[
e^{\displaystyle -i\eta_{r,0}(\bbox{p}_{f,r})}-e^{\displaystyle -i
\eta_{r,0}(\underline{\bbox{p}}_{f,r})}\right]
\enspace .
\label{N0}
\end{eqnarray}

With relations just mentioned above the matrix boundary condition (\ref{zai1})
reduces only to a single scalar equation. In the particular case of this
section one should not forget either that equations (\ref{symmetry}),
(\ref{N0}) are valid only for sufficiently small $\alpha$. For these small
values of $\alpha$ the estimation $\left|\eta_{r,0}(\bbox{p}_{f,r})
-\eta_{r,0}(\underline{\bbox{p}}_{f,r})\right|\propto \alpha$
holds and the quantity
$\alpha g_{r,0} (\bbox{ p}_{f,r},\varepsilon_B(\bbox{p}_{f,l}))$
is sufficiently small and vanishes in the limit $\alpha\to 0$.
This can be seen, in particular, from (\ref{N0}),
since in the limit of impenetrable boundary amplitudes
${f}_{N0}(\bbox{p}_{f,l},\varepsilon_B(\bbox{p}_{f,l}))$\ ,
${f}^+_{N0}(\underline{\bbox{p}}_{f,l},\varepsilon_B(\bbox{p}_{f,l}))$
should vanish. For sufficiently small transmission coefficient
and therefore small $\left[\eta_{r,0}(\bbox{p}_{f,r})
-\eta_{r,0}(\underline{\bbox{p}}_{f,r})\right]$ a remarkably simple boundary
condition follows from (\ref{zai1})
\begin{equation}
\eta_{r,0}(\bbox{p}_{f,r})
-\eta_{r,0}(\underline{\bbox{p}}_{f,r})=-2i\alpha\approx-iD(\bbox{p}_{f})
\enspace .
\label{bcnis}
\end{equation}

The complex pole $\varepsilon_B(\bbox{p}_{f,l})$ can be
found now on the basis of equations (\ref{pe}), (\ref{asr}), (\ref{bcnis}).
For broadened zero energy bound states,
one should assume $\phi_r(\bbox{p}_{f,r})=0$,\
$\phi_r(\underline{\bbox{p}}_{f,r})=\pi$  ensuring the existence of
that state in the limit $D\to 0$ for the trajectory.
Then the solution of Eq.(\ref{pe}) for the superconductor
coincides with (\ref{linsor}) in the absence of a supercurrent across
the junction:
$\delta\eta_{r}(\bbox{p}_{f,r},x)=
-\varepsilon_B(\bbox{p}_{f,r})
{\rm sgn}\left({\rm v}^r_{f,x}(\bbox{p}_{f,r})\right)/
|\tilde{\Delta}(\bbox{p}_{f,r},x)|$. The definition of
$|\tilde{\Delta}(\bbox{p}_{f,r},x)|=|\tilde{\Delta}^r(\bbox{p}_{f,r},x)|$
is given in (\ref{effder}).
Substituting this solution into
boundary condition (\ref{bcnis}) one gets a purely imaginary value
$\varepsilon_B(\bbox{p}_{f,r})=-i\Gamma(\bbox{p}_{f,r})$ for the
pole :
\begin{equation}
\Gamma(\bbox{p}_{f,r})= D(\bbox{p}_{f})\frac{
\displaystyle |\tilde{\Delta}(\bbox{p}_{f,r},0)|
|\tilde{\Delta}(\underline{\bbox{p}}_{f,r},0)|}{\displaystyle
|\tilde{\Delta}(\bbox{p}_{f,r},0)|+|\tilde{\Delta}(\underline{
\bbox{p}}_{f,r},0)|}
\enspace .
\label{br}
\end{equation}
According to (\ref{br}), the effects of surface pair breaking on the broadening
of the zero energy bound state is taken into account by introducing effective
order parameter values (\ref{effder}) taken at the interface. This effective
order parameter reduces into order parameter of the superconductor when
neglecting surface pair breaking. This is entirely analogous to what was
obtained in Sec. III.A for the effects of surface pair breaking on the energies
of low energy bound states. In the particular case of no surface pair breaking
Eq.(\ref{br}) coincides with the result obtained in Ref.\ \onlinecite{walker}
for crystal orientations when $|\Delta(\bbox{p}_{f,r})|=|\Delta(
\underline{\bbox{p}}_{f,r})|$.

\section*{Acknowledgments}

I should like to thank Juhani Kurkij\"arvi for kind hospitality during my stay
in \AA bo Akademi, for helpful comments and critical reading of the manuscript.
I thank W.~Belzig, M.~Fogelstr\"om, F.~Wilhelm and A.~Zaikin for discussions.
This work was supported by the Academy of Finland, research Grant No. 4385. The
financial support of Russian Foundation for Basic Research under grant
No.~99-02-17906 and grant INTAS-RFBR-95-1305 is acknowledged.


\section*{Appendix}

One can derive from the boundary conditions (\ref{zai2}), (\ref{red}) with the
substitution (\ref{symmetry}), the following relations among
interface values of the propagator $\tilde g$ taken both for
incident and (or) for outgoing momenta:
$$
{\tilde g}_{l,0}(\bbox{p}_{f,l})\sin\left(\frac{\displaystyle
\eta_{l,0}(\bbox{p}_{f,l})-
\eta_{r,0}(\bbox{p}_{f,r})}{\displaystyle 2}\right)
\sin\left(\frac{\displaystyle \eta_{l,0}(\bbox{p}_{f,l})-
\eta_{r,0}(\underline{\bbox{p}}_{f,r})}{\displaystyle 2}\right)=
\qquad\qquad\qquad\qquad\qquad
$$
\begin{equation}
\quad\qquad\qquad\qquad\qquad\quad
{\tilde g}_{l,0}(\underline{\bbox{p}}_{f,l})
\sin\left(\frac{\displaystyle \eta_{l,0}(\underline{\bbox{p}}_{f,l})-
\eta_{r,0}(\bbox{p}_{f,r})}{\displaystyle 2}\right)
\sin\left(\frac{\displaystyle \eta_{l,0}(\underline{\bbox{p}}_{f,l})-
\eta_{r,0}(\underline{\bbox{p}}_{f,r})}{\displaystyle 2}\right)
\enspace ,
\label{ll}
\end{equation}

$$
{\tilde g}_{r,0}(\bbox{p}_{f,r})
\sin\left(\frac{\displaystyle \eta_{r,0}(\bbox{p}_{f,r})-
\eta_{l,0}(\bbox{p}_{f,l})}{\displaystyle 2}\right)
\sin\left(\frac{\displaystyle \eta_{r,0}(\bbox{p}_{f,r})-
\eta_{l,0}(\underline{\bbox{p}}_{f,l})}{\displaystyle 2}\right)=
\qquad\qquad\qquad\qquad\qquad
$$
\begin{equation}
\quad\qquad\qquad\qquad\qquad\quad
{\tilde g}_{r,0}(\underline{\bbox{p}}_{f,r})
\sin\left(\frac{\displaystyle \eta_{r,0}(\underline{\bbox{p}}_{f,r})-
\eta_{l,0}(\bbox{p}_{f,l})}{\displaystyle 2}\right)
\sin\left(\frac{\displaystyle \eta_{r,0}(\underline{\bbox{p}}_{f,r})-
\eta_{l,0}(\underline{\bbox{p}}_{f,l})}{\displaystyle 2}\right)
\enspace ,
\label{rr}
\end{equation}

$$
{\tilde g}_{l,0}(\bbox{p}_{f,l})
\sin\left(\frac{\displaystyle \eta_{l,0}(\bbox{p}_{f,l})-
\eta_{r,0}(\bbox{p}_{f,r})}{\displaystyle 2}\right)
\sin\left(\frac{\displaystyle \eta_{l,0}(\bbox{p}_{f,l})-
\eta_{l,0}(\underline{\bbox{p}}_{f,l})}{\displaystyle 2}\right)=
\qquad\qquad\qquad\qquad\qquad
$$
\begin{equation}
\qquad\qquad\qquad\qquad\qquad
{\tilde g}_{r,0}(\underline{\bbox{p}}_{f,r})
\sin\left(\frac{\displaystyle \eta_{r,0}(\underline{\bbox{p}}_{f,r})-
\eta_{l,0}(\underline{\bbox{p}}_{f,l})}{\displaystyle 2}\right)
\sin\left(\frac{\displaystyle \eta_{r,0}(\underline{\bbox{p}}_{f,r})-
\eta_{r,0}(\bbox{p}_{f,r})}{\displaystyle 2}\right)
\enspace ,
\label{lr}
\end{equation}

$$
{\tilde g}_{r,0}(\bbox{p}_{f,r})
\sin\left(\frac{\displaystyle \eta_{r,0}(\bbox{p}_{f,r})-
\eta_{l,0}(\bbox{p}_{f,l})}{\displaystyle 2}\right)
\sin\left(\frac{\displaystyle \eta_{r,0}(\bbox{p}_{f,r})-
\eta_{r,0}(\underline{\bbox{p}}_{f,r})}{\displaystyle 2}\right)=
\quad\qquad\qquad\qquad\qquad
$$
\begin{equation}
\qquad\qquad\qquad\qquad\qquad\quad
{\tilde g}_{l,0}(\underline{\bbox{p}}_{f,l})
\sin\left(\frac{\displaystyle \eta_{l,0}(\underline{\bbox{p}}_{f,l})-
\eta_{r,0}(\underline{\bbox{p}}_{f,r})}{\displaystyle 2}\right)
\sin\left(\frac{\displaystyle \eta_{l,0}(\underline{\bbox{p}}_{f,l})-
\eta_{l,0}(\bbox{p}_{f,l})}{\displaystyle 2}\right)
\enspace .
\label{rl}
\end{equation}

$$
{\tilde g}_{l,0}(\bbox{p}_{f,l})
\sin\left(\frac{\displaystyle \eta_{l,0}(\bbox{p}_{f,l})-
\eta_{r,0}(\underline{\bbox{p}}_{f,r})}{\displaystyle 2}\right)
\sin\left(\frac{\displaystyle \eta_{l,0}(\bbox{p}_{f,l})-
\eta_{l,0}(\underline{\bbox{p}}_{f,l})}{\displaystyle 2}\right)=
\qquad\qquad\qquad\qquad\qquad
$$
\begin{equation}
\quad\qquad\qquad\qquad\qquad
-{\tilde g}_{r,0}(\bbox{p}_{f,r})
\sin\left(\frac{\displaystyle \eta_{r,0}(\bbox{p}_{f,r})-
\eta_{l,0}(\underline{\bbox{p}}_{f,l})}{\displaystyle 2}\right)
\sin\left(\frac{\displaystyle \eta_{r,0}(\bbox{p}_{f,r})-
\eta_{r,0}(\underline{\bbox{p}}_{f,r})}{\displaystyle 2}\right)
\enspace ,
\label{lrp}
\end{equation}

$$
{\tilde g}_{l,0}(\underline{\bbox{p}}_{f,l})
\sin\left(\frac{\displaystyle \eta_{l,0}(\underline{\bbox{p}}_{f,l})-
\eta_{r,0}(\bbox{p}_{f,r})}{\displaystyle 2}\right)
\sin\left(\frac{\displaystyle \eta_{l,0}(\underline{\bbox{p}}_{f,l})-
\eta_{l,0}(\bbox{p}_{f,l})}{\displaystyle 2}\right)=
\qquad\qquad\qquad\qquad\qquad
$$
\begin{equation}
\qquad\qquad\qquad\qquad\quad
-{\tilde g}_{r,0}(\underline{\bbox{p}}_{f,r})
\sin\left(\frac{\displaystyle \eta_{r,0}(\underline{\bbox{p}}_{f,r})-
\eta_{l,0}(\bbox{p}_{f,l})}{\displaystyle 2}\right)
\sin\left(\frac{\displaystyle \eta_{r,0}(\underline{\bbox{p}}_{f,r})-
\eta_{r,0}(\bbox{p}_{f,r})}{\displaystyle 2}\right)
\enspace .
\label{rlp}
\end{equation}




\begin{references}
\bibitem{sns}
Ansatz (\ref{symmetry}) works well inside a normal metal layer if
Andreev bound states arise there. For instance, in a clean SNS system with
fully transparent interfaces  one can easily get conventional Andreev descrete
quasiparticle subgap levels basing on the equation (\ref{pe}) and
asymptotic conditions (\ref{asr}), (\ref{asl}).
\bibitem{ans}
In the normal metal halfspace one of two Andreev amplitudes $u(\bbox{p}_f,x)$,
$v(\bbox{p}_f,x)$ with a given $\bbox{p}_f$ vanishes. The other one
is finite on account of the proximity effect and decreases exponentially toward
the bulk normal metal. In this special case Eq.(\ref{and}) is not suitable for
describing the Andreev amplitudes.
\end{references}
\end{document}